%% file: linear_rx_capacity_largeN.tex
\documentclass[conference]{IEEEtran}
\IEEEoverridecommandlockouts

\usepackage{amsmath}
\usepackage{amssymb}
\usepackage{amsfonts}
\usepackage{epsfig}
\usepackage{calc}
\usepackage{subfigure}

\usepackage[usenames]{color}
\usepackage[normalem]{ulem}

\input{macros}

\newcommand{\beq}{\begin{equation}}
\newcommand{\eeq}{\end{equation}}

\newcommand{\bea}{\begin{eqnarray}}
\newcommand{\eea}{\end{eqnarray}}
\newcommand{\bean}{\begin{eqnarray*}}
\newcommand{\eean}{\end{eqnarray*}}

\newcommand{\bit}{\begin{itemize}}
\newcommand{\eit}{\end{itemize}}
\newcommand{\ben}{\begin{enumerate}}
\newcommand{\een}{\end{enumerate}}

\title{Performance of MMSE MIMO Receivers:\\ A Large $N$ Analysis for Correlated Channels}

\author{\IEEEauthorblockN{Aris L. Moustakas}
\IEEEauthorblockA{Dept. of Physics\\
National \& Capodistrian University of Athens\\
157 84  Athens, Greece\\
arislm@phys.uoa.gr
\thanks{This research was supported in part by the project ``Newcom++'' with No. EU-IST-NoE-FP6-2007-216715.}
} %
\and %
\IEEEauthorblockN{K. Raj Kumar and Giuseppe Caire}
\IEEEauthorblockA{Dept. of EE - Systems \\
University of Southern California\\
Los Angeles, CA 90007, USA\\
\{rkkrishn,caire\}@usc.edu}%
} %

\def\BibTeX{{\rm B\kern-.05em{\sc i\kern-.025em b}\kern-.08em
    T\kern-.1667em\lower.7ex\hbox{E}\kern-.125emX}}

\begin{document}
\maketitle

\begin{abstract}
Linear receivers are considered as an attractive low-complexity alternative to optimal processing for multi-antenna MIMO communications. In this paper we characterize the performance of MMSE MIMO receivers in the limit of large antenna numbers in the presence of channel correlations. Using the replica method, we generalize our results obtained in \cite{Kumar2008_DMT} to Kronecker-product correlated channels and calculate the asymptotic mean and variance of the mutual information of a MIMO system of parallel MMSE subchannels. The replica method allows us to use the ties between the optimal receiver mutual information and the MMSE SIR of Gaussian inputs to calculate the joint moments of the SIRs of the MMSE subchannels. Using the methodology discussed in \cite{Kumar2008_DMT} it can be shown that the mutual information converges in distribution to a Gaussian random variable. Our results agree very well with simulations even with a moderate number of antennas.
\end{abstract}


\section{Introduction}
\label{sec:intro}

The next generation of wireless communication systems
is expected to capitalize on the large gains in spectral
efficiency and reliability promised by MIMO multi-antenna
communications \cite{Tel, Foschini1998_BLAST1} and include MIMO
technology as a fundamental component of their physical layer
\cite{ieee802.11n}.
To keep the complexity within limits, the standardization and the current commercial implementation
of these systems has focused on schemes, based on linear spatial equalization.

\begin{figure}[ht]
\begin{center}
\includegraphics[width=8.7cm]{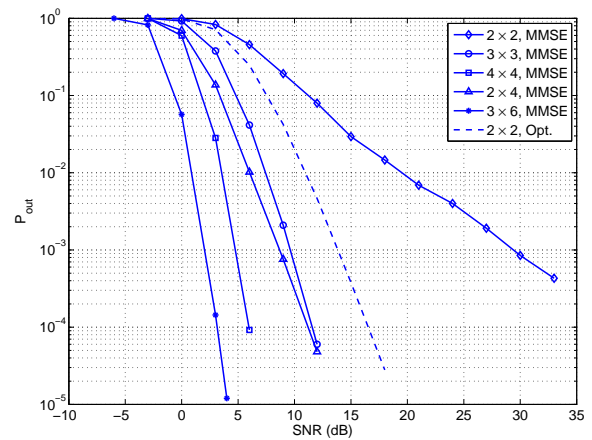}
\caption{Comparing the outage probability of optimal and MMSE
receivers, $R = 3$ bpcu.\label{fig:increase_antennas}}
\end{center}
\end{figure}

Consider the following system
design issue: for a given target spectral efficiency, block-error
rate, operating SNR, and receiver computational complexity
(including power consumption, VLSI chip area etc.) how many antennas
do we need at the transmitter and receiver? Consider the outage
probability curves of Fig.~\ref{fig:increase_antennas} and suppose
that we wish to achieve a rate of $R = 3$ bpcu with block-error rate
of $10^{-3}$ at SNR not larger than 15 dB. With $M = N = 2$ antennas
this target performance is achieved by an optimal receiver, but is
not achieved by the MMSE receiver. However, with $M = 2, N= 4$ or $M
= N = 3$ the target performance is achieved also by the MMSE
receiver. It turns out that, in some cases, adding antennas may be
more convenient than insisting on high-complexity receiver
processing.

It is therefore interesting to analyze the outage probability of
a linear  receiver with coding across the
antennas in the regime of fixed SNR $\rho$ and rate $R$. This
analysis is difficult due to the fact that, for finite $M, N$, the
joint distribution of the channel SINRs $\{\gamma_k\}$ in
(\ref{sinr-mmse}) escapes a closed-form expression. This problem
can be overcome by considering the system in the limit of a large number of antennas.

In this paper, we generalize the approach developed in \cite{Kumar2008_DMT} to handle channels with correlations across the antennas. We will sketch the proof of the distribution of the mutual information approaching a gaussian distribution in the large antenna limit and we will calculate, applying the random matrix theory and the replica approach to characterize the limiting joint Gaussian distribution of the SINRs for the case of correlated channels and obtain the statistics of the mutual information of linear MMSE receivers. The analysis provides accurate results even for a moderate number of antennas and allows to quantify how the performance loss in terms of diversity suffered by linear receivers may be recovered by increasing the number of antennas. This analysis prompts to the conclusion that  in order to achieve a desired target spectral efficiency and block-error rate,
at given SNR and receiver complexity, increasing the number of antennas and using simple
linear receiver processing may be, in fact, a good design option.

It should be pointed out that asymptotic Gaussianity has been shown \cite{Moustakas2003_MIMO1,
Hachem2007_NewApproachGaussianMIMO} for the MIMO channel mutual
information given by the ``log-det'' formula, whose cumulative
distribution function (cdf) yields the block-error rate achievable
by optimal coding and decoding. At the same time
the asymptotic Gaussianity of the SINR of a single MMSE receiver channel
was derived in \cite{Tse2000_MMSEFluctuations, Liang2007_MMSEAsymptotics}, without looking
at the {\em joint Gaussianity} of all SINRs for all these channels.
The joint Gaussianity of SINRs  is crucial in the analysis of the statistics of the total mutual information.

\section{System model} \label{system-model}

The output of the underlying frequency flat slowly-varying MIMO channel is given by
\begin{equation} \label{eq:mimo}
\yv_t = \Hm \xv_t + \wv_t, \;\;\; t = 1,\ldots,T,
\end{equation}
where $\xv_t \in \CC^M$ denotes the channel input vector at
channel use $t$, $\wv_t \sim \Cc\Nc(\zerov, N_0 \Id)$ is the
additive spatially and temporally white Gaussian noise and $\Hm
\in \CC^{N \times M}$ is the channel matrix. In this work we make
the assumption that the entries of $\Hm$ are zero-mean Gaussian with separable correlations, i.e.
\begin{equation} \label{eq:mimo_corrs}
\EE\left[H_{ia}H^*_{jb}\right] = R_{ij} T_{ab}
\end{equation}
where $\Rm$ and $\Tm$ are the $N\times N$ and $M\times M$ correlation matrices at the receiver and transmitter, respectively. Also $\Hm$ is assumed constant over the
duration $T$ of a codeword (quasi-static Rayleigh i.i.d. fading
\cite{Tel}). The input is subject to the total power
constraint $\frac{1}{MT} \EE \left[ \|\Xm\|_F^2 \right] \leq E_s$,
where $\Xm = [\xv_1,\ldots,\xv_T]$ denotes a space-time codeword,
uniformly distributed over the space-time codebook $\Xc$, and
$\|\cdot\|_F$ denotes the Frobenius norm. Furthermore, we
define the transmit SNR $\rho$ as the total transmit energy per
time-slot over the noise power spectral density, i.e., $\rho = M
E_s/N_0$. We assume no Channel State Information (CSI) at the transmitter and without loss of generality
we also assume perfect CSI at the receiver. In addition, we make the convenient assumption of Gaussian inputs and large block length $T$.

We consider a MMSE {\em memoryless} receiver with signal to interference and noise ratio at the $k$-th receiver output given by  \cite{Ver}
\begin{eqnarray} \label{sinr-mmse}
\gamma_k & = & \frac{\rho}{M} \hv_k^\herm \left [ \Id + \frac{\rho}{M} \Hm_k \Hm_k^\herm \right ]^{-1} \hv_k \nonumber \\ \label{sinr-mmse1}
&=& \frac{1}{\left [\left (\Id + \frac{\rho}{M} \Hm^\herm \Hm
\right )^{-1} \right ]_{kk}} - 1,
\end{eqnarray}
where $\Hm_k$ denotes the $N \times (M-1)$ matrix obtained by
removing the $k^{\text{th}}$ column, $\hv_k$, from $\Hm$. Further assuming ideal interleaving and coding across antennas, the performance metric characterizing such a system is the outage probability $P_{out}(R)=Prob(I_N\leq R)$ where the mutual information $I_N$ is
\begin{equation} \label{IN}
I_N \triangleq \sum_{k=1}^M \log \left ( 1 + \gamma_k \right )
\end{equation}

\section{Methodology} \label{sec:MMSE_MI_distrib}

In the next subsection we will sketch the methodology used to show
the asymptotic Gaussianity of the mutual information.
Subsequently, in Section~\ref{sec:Joint_cum_moments12} we will
calculate the first and second cumulant moments of the MMSE SINR for correlated channels, which suffice to characterize the mutual information limiting distribution.

\subsection{Asymptotic Gaussianity of the mutual information}
\label{sec:gaussianity_ofMI}

In order to prove the asymptotic Gaussianity of the mutual information $I_N$,
we will need to analyze its cumulant moments ${\cal C}_n$ for large $N$, given by the expansion of the logarithm of its characteristic function $\Phi_N(\omega) \triangleq \EE \left[ e^{-j \omega I_N} \right]$,
\begin{equation}\label{eq:cum_mg_function_def}
\log \Phi_N(\omega) = \sum_{n=1}^\infty
\frac{(-\mu)^n}{n!} {\cal C}_n,
\end{equation}
For example, the first two cumulant moments of $X$ are the mean ${\cal C}_1 = \EE[X]$ and the variance ${\cal C}_2 = \text{Var}[X]$ of $I_N$. When all cumulant moments ${\cal C}_n$ for $n>2$ vanish in the limit $N\rightarrow \infty$ it can be shown\cite{Kumar2008_DMT} that the distribution of $I_N$ approaches a Gaussian.

In Section~\ref{sec:Joint_cum_moments12}, we will show that in the
limit of large $N$ and $M = \beta N$ with $\beta \leq 1$,
\begin{eqnarray}\label{eq:asympt_C_1}
{\cal C}_1 & = & m_1 + o(1)  \\
\label{eq:asympt_C_2}
{\cal C}_2 & = & \sigma^2 + o(1)
\end{eqnarray}
where $m_1 = M c_{10} + c_{11}$, and where $c_{10}$, $c_{11}$ and
$\sigma^2$ are constants independent of $N$ for which we give general
analytic expressions and for i.i.d. channels closed form expressions. In \cite{Kumar2008_DMT, Moustakas2003_MIMO1}
we show that all higher-order cumulants of the mutual information
asymptotically vanish for large $N$.

\subsection{Joint cumulant moments of the SINRs of order 1 and 2}
\label{sec:Joint_cum_moments12}

Our goal is to calculate the mean and variance of $I_N$. Since $I_N$
consists of a sum of mutual informations of the virtual channels
(see (\ref{IN})), the building blocks of the cumulant moments of $I_N$ are the
joint cumulant moments of the SINRs $\{\gamma_k\}$. It can be shown\cite{Kumar2008_DMT} that to calculate the mean $\EE[I_N]$ to order $O(1)$ we need the mean of $\gamma_k$ to order $1/N$ and the leading term in the variance of $\gamma_k$. In addition, the variance of $I_N$ will require the cumulant joint cross-correlations $\EE_c[\gamma_k;\gamma_l$.

With the above in mind, we will apply the methodology developed for the calculation of the moments of the ``logdet'' formula (i.e. the optimal receiver mutual information of MIMO correlated Gaussian channels to the analysis of the joint moments of the MMSE SIR's, $\gamma_k$. The link between the two is given by the simple relation \cite{Guo2005_MutualInformation_MMSE}
\begin{eqnarray}\label{eq:MI_MMSESIR_def}
    \frac{d}{dx} \left. Tr\log\left[\Id + \frac{\rho}{M}\Hm \Jm_k(x)\Hm^\herm\right]\right|_{x=0} \\
    Tr \left\{\left[\Id + \frac{\rho}{M}\Hm \Jm_k(0)\Hm^\herm\right]^{-1} \frac{\rho}{M}\Hm \Jm'_k(0)\Hm^\herm \right\}
\end{eqnarray}
$\Jm_k(x)=\Id +(x-1) \deltav_k$, where $\deltav_k$ is a matrix a single non-zero element $(\deltav_k)_{ij} = \delta_{ik}\delta_{jk}$, the RHS of the above equation is exactly $\gamma_k$. For simplicity we will suppress  the $x$-dependence in $\Jm_k(x)$ when not necessary. Thus
\begin{eqnarray}\label{eq:MI_MMSESIR_def2}
    \gamma_k &=& Tr \left\{\left[\Id + \frac{\rho}{M}\Hm \Jm_k(0)\Hm^\herm\right]^{-1} \frac{\rho}{M}\Hm \deltav_k\Hm^\herm \right\} \\ \nonumber
    &=& \frac{d}{dx}\left. I_N(\Jm_k)\right|_{x=0}
\end{eqnarray}
where we define the $k$'th channel mutual information by $A(\Jm_k(x))=A_k = Tr\left\{\log\left[\Id+\frac{\rho}{M}\Hm\Jm_k(x)\Hm^\herm\right]\right\}$. The above relation allows us to evaluate the joint cumulant moments of $\gamma_k$ through the cumulant moments of $A_k $, e.g.
\begin{eqnarray}\label{eq:MeanMMSESIR}
    \EE_c[\gamma_k] &=& \frac{d}{dx}\left.\EE_c\left[ A_k \right]\right|_{x=0}
    \\ \label{eq:VarMMSESIR}
    \EE_c[\gamma_k;\gamma_l] &=& \left.\frac{d^2\EE_c\left[ A(\Jm_k(x_k))A(\Jm_l(x_l))\right]}{dx_kdx_l}\right|_{x_k=x_l=0}
\end{eqnarray}
The mean of the mutual information in (\ref{eq:MeanMMSESIR}) can be evaluated in many different ways in the asymptotic $N$ limit. We will quote the result below directly \cite{Moustakas2003_MIMO1}.
\begin{eqnarray}
    \label{eq:MeanMI_Ak}
    \EE[A_k] &=& Tr\log\left(\Id+\sqrt{\rho} t_k {\tilde \Tm}_k\right) -Mt_kr_k \\ \nonumber
    &+& Tr\log\left(\Id+\sqrt{\rho} r_k \Rm \right) +O(1/N) \\
    \label{eq:t_MeanMI_Ak}
    t_k &=& \frac{1}{M} Tr\left[\frac{\sqrt{\rho}\Rm}{\Id+\sqrt{\rho}r_k\Rm}\right]\\
    \label{eq:r_MeanMI_Ak}
    r_k &=& \frac{1}{M} Tr\left[\frac{\sqrt{\rho}{\tilde \Tm}_k }{\Id+\sqrt{\rho}t_k{\tilde \Tm}_k }\right]\\
    \label{eq:T_k_tilde_MeanMI_Ak}
    {\tilde \Tm}_k &=& \Tm \Jm_k(x)
\end{eqnarray}
To evaluate  the mean SINR $\EE[\gamma_k]$ we take  the derivative with respect to $x$ of the above quantity and evaluating (\ref{eq:MeanMMSESIR}) we find that the mean SINR $\EE[\gamma_k]$ is equal to
\begin{eqnarray}\label{eq:MeanMMSESIR_largeN}
    \bar{\gamma}_k &=& Tr \left[\frac{t_k\sqrt{\rho}\Tm}{\Id+\sqrt{\rho}t_k\Tm\Jm_k(0)}\deltav_k\right] = \frac{1}{\eta_k}-1
\end{eqnarray}
where $\eta_k$ is $k$-diagonal component of the matrix $\eta$-transform
\begin{eqnarray}\label{eq:eta_k}
    \eta_k &=& \left[\left(\Id+t_k\Tm\sqrt{\rho}\right)^{-1}\right]_{kk}
\end{eqnarray}
The relation with the $\eta_k$ transform \cite{Tulino2004_RMTInfoTheoryReview} can be readily seen by comparing (\ref{eq:eta_k}) with (\ref{sinr-mmse1}). The quantity $t_k$ in the above result is evaluated by solving (\ref{eq:t_MeanMI_Ak}), (\ref{eq:r_MeanMI_Ak}) with $\Jm_k(0)$. The leading order result in $N$ can be obtained by making the approximation $\Jm_k=\Id$ in (\ref{eq:r_MeanMI_Ak}). We call these solutions $t_0$, $r_0$. To get the $1/N$ correction to $\gamma_k$ we need to evaluate the correction  $\delta t=t-t_0$ due to the fact that $\Jm_k(0) = \Id-\deltav_k$. Solving a pair of coupled linear equations for $\delta t$ and the corresponding correction $\delta r$ and inserting the result in (\ref{eq:MeanMMSESIR_largeN}) we get
\begin{eqnarray}\label{eq:MeanMMSESIR_largeN_1}
    \delta\bar{\gamma}_k = \frac{1}{M} \frac{\eta_k'^2}{\eta_k^3}\frac{M_{r2}}{1-M_{t2}M_{r2}}
\end{eqnarray}
where
\begin{eqnarray}
    \label{eq:eta_k_prime}
    \eta_k' &=& \frac{d\eta_k}{dt_k} = -\left[\frac{\Tm\sqrt{\rho}}{\left(\Id+t_k\Tm\sqrt{\rho}\right)^2}\right]_{kk} \\
    \label{eq:Mt2}
    M_{t2} &=& \frac{1}{M} Tr\left[\left(\frac{\Tm\sqrt{\rho}}{\Id+t_k\Tm\sqrt{\rho}}\right)^2\right] \\
    \label{eq:Mr2}
    M_{r2} &=& \frac{1}{M} Tr\left[\left(\frac{\Rm\sqrt{\rho}}{\Id+r_k\Rm\sqrt{\rho}}\right)^2\right]
\end{eqnarray}
To evaluate the second {\em joint} moments of the mutual information as in (\ref{eq:VarMMSESIR}), we can take  derivatives of the joint moment generating function, defined by
\begin{eqnarray}\label{eq:JCMGenFunI_N_def}
    g_{kl}(\mu,\nu)= \log\EE\left[ e^{-\mu A_k(\Jm_k)-\nu A_l(\Jm_l)}\right]
\end{eqnarray}
evaluated at $\mu=\nu=0$. Thus,
\begin{eqnarray}
    \label{eq:VarMMSESIR_2}
    \EE_c[\gamma_k;\gamma_l] = \left.\frac{d^2}{dx_kdx_l}\frac{d^2g_{kl}(\mu,\nu)}{d\mu d\nu}\right|_{x_k=x_l=0; \mu=\nu=0}
\end{eqnarray}

We will leave the details for the appendix. Here we will just quote the result:
\begin{eqnarray}
    \label{eq:VarMMSESIR_3}
\EE_c\left[ A_k;A_l\right] &=& -\log\left[1-M_{t_k,t_l}M_{r_k,r_l}\right]
\end{eqnarray}
\begin{eqnarray}
    \label{eq:Mt1t2}
M_{t_k,t_l} &=& \frac{\rho}{M}Tr\left[\frac{\Jm_k\Tm}{\Id+t_k\Jm_k\Tm\sqrt{\rho}}
\frac{\Jm_l\Tm}{\Id+t_l\Jm_l\Tm\sqrt{\rho}}\right]\\
    \label{eq:Mr1r2}
M_{r_k,r_l} &=& \frac{\rho}{M}Tr\left[\frac{\Rm}{\Id+\Rm r_k\sqrt{\rho}}\frac{\Rm}{\Id+\Rm r_l\sqrt{\rho}}\right]
\end{eqnarray}
where $t_k, t_l, r_k, r_l$ are solutions to the equations (\ref{eq:t_MeanMI_Ak}), (\ref{eq:r_MeanMI_Ak}) for $\Jm_k$ and $\Jm_l$ respectively. Taking now derivatives with respect to $x_k$, $x_l$ and setting these to  zero we arrive at the following robust expression for the joint moments of $\gamma_i$, $\gamma_j$
\begin{eqnarray}\label{eq:gamma_gen_corr_matrix}
\Sigma_{i,j} = \EE_c[\gamma_i;\gamma_j] = \left.-\frac{d^2\log\left[1-M_{t_i,t_j}M_{r_i,r_j}\right] }{dx_idx_j} \right|_{x_i=x_j=0}
\end{eqnarray}
where the $x_i$ derivatives should also be applied on $t_i$, $r_i$, whose $x$ dependence can be traced to
(\ref{eq:t_MeanMI_Ak}), (\ref{eq:r_MeanMI_Ak}). When $i=j$, the matrices $\deltav_i=\deltav_j$, but the derivatives over $x_i$ and $x_j$ should be taken separately. The matrix $\Sigmam$ is the correlation matrix of the SIR's $\gamma_k$. In the limit of uncorrelated channels $\Hm$, i.e. $\Rm=\Id$ and $\Tm=\Id$, all diagonal and off-diagonal elements of $\Sigmam$ are equal with each other. The result can be written in a very simple form in terms of the mean $\gamma_k$, $\EE[\gamma_k]=g=t_k\sqrt{\rho}=(\rho(1-\beta)-\beta+\sqrt{(\rho(1-\beta)-\beta)^2+4\rho\beta})/(2\beta)$ \cite{Kumar2008_DMT} as follows:
\begin{eqnarray}\label{eq:gamma_iid_corr_matrix_diag}
\Sigma_{1,1} &=& \frac{1}{M} \frac{\beta g^2}{1-\beta\frac{g^2}{(1+g)^2}} \equiv \frac{v_d}{M} \\
\label{eq:gamma_iid_corr_matrix_offdiag}
\Sigma_{1,2} &=& \frac{\beta^2 g^3\left[g\left(1-\beta\frac{g^2}{(1+g)^2}\right)-2\right]}{M^2(1+g)^2\left(1-\beta\frac{g^2}{(1+g)^2}\right)^3} \\ \nonumber
&+& \frac{\beta^3 g^4}{M^2(1+g)^4\left(1-\beta\frac{g^2}{(1+g)^2}\right)^4} \equiv \frac{v_{od}}{M^2}
\end{eqnarray}
As we see, the diagonal correlations of $\gamma_k$ are $O(1/M)$, while the off-diagonal ones are $O(1/M^2)$. This scaling holds also for the general case (\ref{eq:gamma_gen_corr_matrix}). Despite the different appearance, the above result is identical to the one presented in \cite{Kumar2008_DMT}.

\subsection{Gaussian approximation and outage probability}

We will now use the previous results to give an
explicit Gaussian approximation for the outage probability of the
MMSE receiver with coding across the antennas in the regime
of fixed SNR and large number of antennas.

We start with the mean $\EE[I_N]$. Taking into account the fact that the SIR's $\gamma_k$ are asymptotically jointly gaussian with correlations given by (\ref{eq:gamma_gen_corr_matrix}), we may expand (\ref{IN}) in powers of $1/M$ as follows
\begin{eqnarray}\label{eq:MI_MMSE_mean}
  {\cal C}_1 &=& \EE[I_N] = \sum_{k=1}^M \EE[ \log(1+\gamma_k)]  \\ \nonumber
  &=& \sum_k \log\left(1+\bar{\gamma}_k\right) + \sum_k \left(\delta \bar{\gamma}_k + \Sigma_{kk}\right)
\end{eqnarray}
where $\bar{\gamma}_k$, $\delta\bar{\gamma}_k$ and $\Sigma_{kk}$ are given by (\ref{eq:MeanMMSESIR_largeN}), (\ref{eq:MeanMMSESIR_largeN_1}) and (\ref{eq:gamma_gen_corr_matrix}), respectively. The first sum above is equal to $Mc_{10}$ and gives the leading order term of the mean of the mutual information $I_N$. The second sum, which is $O(1)$, since both summands are $O(1/M)$, is $c_{11}$ appearing in $m_1$ in (\ref{eq:asympt_C_1}). While, the coefficient $c_{10}$ is known, the evaluation of the correction $c_{11}$ is novel, to the best of our knowledge.

We now turn to the variance of the mutual information. Generalizing the approach used in \cite{Kumar2008_DMT} we can show that the variance can be written as
\begin{eqnarray}\label{eq:MI_MMSE_sigma1}
{\cal C}_2 &=& \sum_k \frac{\Sigma_{kk}}{(1+\bar{\gamma}_k)^2} + \sum_{k\neq l} \frac{\Sigma_{kl}}{(1+\bar{\gamma}_k)(1+\bar{\gamma}_l)} \\
&=& \frac{v_d +
v_{od}}{(1+g)^2} + o(1)
\end{eqnarray}
The second line is valid when the channel has no correlations, with $v_{d}$, $v_{od}$ given by (\ref{eq:gamma_iid_corr_matrix_diag}), (\ref{eq:gamma_iid_corr_matrix_offdiag}), respectively. The evaluation of ${\cal C}_2=\sigma$ is novel, to the best of our knowledge.

Under this Gaussian approximation, we can easily evaluate the outage probability
for fixed SNR, $\beta$ and number of antennas $M$ as follows:
\begin{equation}\label{eq:gaussian_approx_mmse_MI}
P_{\rm out}(R) \approx Q \left ( \frac{R - {\cal C}_1}{\sqrt{{\cal C}_2}} \right )
\end{equation}
where $Q(x) = \int_x^\infty \frac{1}{\sqrt{2\pi}} e^{-t^2/2} dt$ is the Gaussian tail function.

\subsection{Simulations and comparisons}

\begin{figure}[ht]
\begin{center}
\hspace*{-0.2in}
\includegraphics[width=10.3cm]{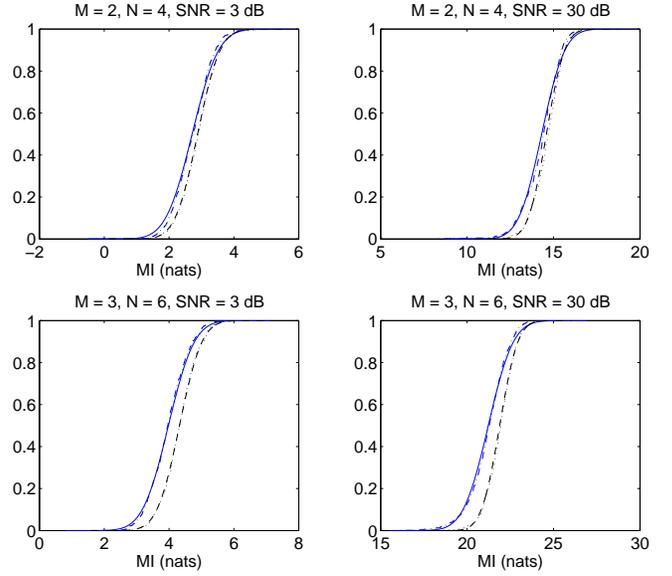}
\caption{CDF of the mutual information (MI) for the MMSE and optimal
receivers, for $M = 2,3$, $\beta = 0.5$ and $\rho = 3, 30$ dB. The
solid blue line is the analytical result for the MMSE, the dot-dash
blue is MMSE empirical, the dashed black is Optimal receiver
analytical and the dotted black is the optimal receiver empirical.
\label{fig:CDF_MI_2}}
\end{center}
\end{figure}

In this section we validate the asymptotic analysis by
comparing the Gaussian approximation to the outage
probability with finite-dimensional Monte Carlo simulation in the case of uncorrelated channels.
For the sake of comparison, we also consider the outage probability
of the optimal receiver, given by the log-det cdf $P( I_N^{\sf opt}
\leq R)$, where $I_N^{\sf opt} = \log \det (\Id + \frac{\rho}{M}
\Hm\Hm^\herm)$.  Using the results for the mean and the variance, we plot the CDF of
the (Gaussian) mutual information for the MMSE and optimal receiver
in Figs.~\ref{fig:CDF_MI_2},\ref{fig:CDF_MI}. Both analytical and
empirical results are plotted, for a wide range of $M,N$ and SNRs.
We notice that the analytical and empirical results match closely,
for even moderate number of antennas and not too large SNRs.

\section{Conclusions} \label{conclusions}

Novel wireless communication systems are targeting very large spectral
efficiencies and will operate at high SNR thanks to hot-spots and pico-cell arrangements.
System design is targeting even higher data rates, up to 1Gb/s, in 40 MHz
system bandwidth. For such systems, the use of low-complexity
linear receivers in a {\em separated detection and decoding}
architecture as those examined in this paper may be
mandatory because of complexity and power consumption.

In this paper we investigated the asymptotic performance of the linear MMSE receiver with coding across the antennas
in the regime of fixed SNR and large (but finite) number of
antennas for a general Kronecker-correlated Gaussian channel. We showed that the corresponding mutual information
has statistical fluctuations that converge in distribution to a Gaussian random variable, and we
computed its mean and variance analytically. This paper used a different method in calculating the moments of the mutual information, namely the replica method, but found identical results with previous approaches \cite{Kumar2008_DMT}.

Based on the analysis carried out in this work, we may summarize
some considerations on system design. In order to achieve a
required target spectral efficiency at given block-error rate and SNR operating point, an
attractive design option may consists of increasing the number of
antennas (especially at the receiver) and using a low-complexity
linear receiver.

\appendix

We now sketch the basic steps in going from (\ref{eq:JCMGenFunI_N_def}) to (\ref{eq:VarMMSESIR_3}). We borrow heavily from the methodology discussed in \cite{Moustakas2003_MIMO1}. The basic idea behind the so-called replica trick is to evaluate  (\ref{eq:JCMGenFunI_N_def}) for integer $\mu$ and $\nu$ and then analytically continue to continuous $\mu$, $\nu$ values. After introducing the $\mu\times\mu$ matrices ${\cal \Tm}_1$, ${\cal \Rm}_1$, the $\nu\times\nu$ matrices ${\cal \Tm}_2$, ${\cal \Rm}_2$ and the $\mu\times\nu$ matrices ${\cal \Tm}_3$, ${\cal \Rm}_3$, following steps discussed in \cite{Moustakas2003_MIMO1} we can integrate over the channel matrix $\Hm$ and express (\ref{eq:JCMGenFunI_N_def}) as
\begin{eqnarray}
  g(\mu,\nu) = \int D\left({\cal T}_1;{\cal R}_1;{\cal T}_2;{\cal R}_2;{\cal T}_3;{\cal R}_3;\right) e^{-{\cal S}}
\end{eqnarray}
\begin{eqnarray}
  {\cal S} &=& \log\det\left(\Id_M\otimes\Id_\mu + \bar{\Jm}_k\otimes{\cal T}_1 \right) \\ \nonumber
  &+&  \log\det\left(\Id_N\otimes\Id_\mu + \bar{\Rm} \otimes {\cal R}_1 \right) \\ \nonumber
  &-&Tr\left[{\cal T}_1 {\cal R}_1 +{\cal T}_2 {\cal R}_2 +{\cal T}_3 {\cal R}_3^\herm +{\cal T}_3^\herm {\cal R}_3\right] \\ \nonumber
  &+&  \log\det\left(\Id_N\otimes\Id_\nu + \bar{\Rm}\otimes{\cal R}_2  \right. \\ \nonumber
  &-&\left.\bar{\Rm}\otimes {\cal R}_3^\herm \left[\Id_N\otimes\Id_\mu + \bar{\Rm} \otimes {\cal R}_1 \right]^{-1} \Rm\otimes {\cal R}_3 \right) \\ \nonumber
  &+&  \log\det\left(\Id_M\otimes\Id_\nu + \bar{\Jm}_l\otimes{\cal T}_2  - \bar{\Jm}_l^{1/2}\bar{\Jm}_k^{1/2}\otimes {\cal T}_3^\herm \right. \\ \nonumber
  &\cdot&\left.\left[\Id_M\otimes\Id_\mu + \bar{\Jm}_k \otimes {\cal T}_1 \right]^{-1} \bar{\Jm}_k^{1/2}\bar{\Jm}_l^{1/2}\otimes {\cal T}_3 \right)
\end{eqnarray}
where 
\begin{eqnarray}
  \bar{\Jm}_k &=& \sqrt{\frac{\rho}{M}} \Tm^{1/2}\Jm_k(x_k)\Tm^{1/2}\\ \nonumber
  \bar{\Rm} &=& \sqrt{\frac{\rho}{M}} \Rm
\end{eqnarray}
We now need to specify the saddle point behavior of the integral over the ${\cal T}$, ${\cal R}$. Specifically, we set ${\cal T}_3={\cal R}_3 = 0$ and ${\cal T}_1=\sqrt{M}t_k \Id_\mu$, ${\cal R}_1=\sqrt{M}r_k \Id_\mu$, ${\cal T}_2=\sqrt{M}t_l \Id_\nu$, ${\cal R}_2=\sqrt{M}r_l \Id_\nu$. The values of $r_k$, $t_k$ are determined from saddle point equations (\ref{eq:r_MeanMI_Ak}) and (\ref{eq:t_MeanMI_Ak}) and corresponding equations for $r_l$ and $t_l$. To calculate the joint moment (\ref{eq:VarMMSESIR_3}), we need to calculate the fluctuations due to ${\cal T}_3$, ${\cal R}_3$, since these are the $\mu\times\nu$ matrices that connect the mutual informations $A_k$, $A_l$. Indeed, expanding ${\cal S}$ to second order in these matrices and after we integrate over them we get precisely the term
\begin{equation}\label{eq:gmunu_term}
    -\mu\nu\log(1-M_{t_kt_l}M_{r_kr_l})
\end{equation}
which, upon differentiation over $\mu$, $\nu$ gives (\ref{eq:VarMMSESIR_3}).

\begin{figure}[ht]
\begin{center}
\hspace*{-0.2in}
\includegraphics[width=10.3cm]{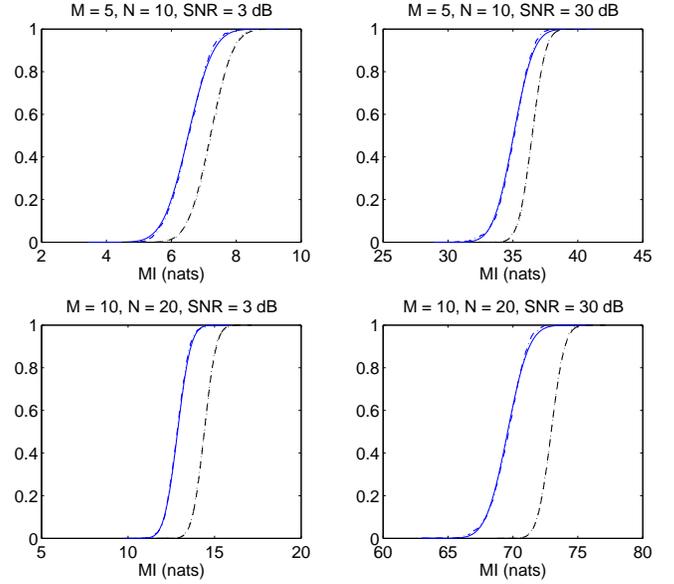}
\caption{CDF of the mutual information (MI) for the MMSE and optimal
receivers, for $M = 5,10$, $\beta = 0.5$ and $\rho = 3, 30$ dB. The
solid blue line is the analytical result for the MMSE, the dot-dash
blue is MMSE empirical, the dashed black is Optimal receiver
analytical and the dotted black is the optimal receiver empirical.
\label{fig:CDF_MI}}
\end{center}
\end{figure}


\end{document}

%% file: macros.tex
\long\def\comment#1{}

\newfont{\bbb}{msbm10 scaled 700}

\newfont{\bb}{msbm10 scaled 1100}
\newcommand{\CC}{\mbox{\bb C}}

\newcommand{\EE}{\mbox{\bb E}}

\newcommand{\hv}{{\bf h}}

\newcommand{\wv}{{\bf w}}

\newcommand{\xv}{{\bf x}}
\newcommand{\yv}{{\bf y}}

\newcommand{\zerov}{{\bf 0}}

\newcommand{\Hm}{{\bf H}}
\newcommand{\Id}{{\bf I}}
\newcommand{\Jm}{{\bf J}}

\newcommand{\Rm}{{\bf R}}

\newcommand{\Tm}{{\bf T}}

\newcommand{\Xm}{{\bf X}}

\newcommand{\Cc}{{\cal C}}

\newcommand{\Nc}{{\cal N}}

\newcommand{\Xc}{{\cal X}}

\newcommand{\deltav}{\hbox{\boldmath$\delta$}}

\newcommand{\Sigmam}{\hbox{\boldmath$\Sigma$}}

\renewcommand{\det}{{\hbox{det}}}

\newcommand{\herm}{{\sf H}}